\documentclass[journal]{IEEEtran}
\IEEEoverridecommandlockouts

\usepackage{cite}
\usepackage{amsmath,amssymb,amsfonts}
\usepackage{algorithmic}
\usepackage{graphicx}
\usepackage{textcomp}
\usepackage{xcolor}
\usepackage{booktabs}
\usepackage{color,soul}
\usepackage{tabularx,hhline}
\usepackage{multirow}
\usepackage{hyperref}
\usepackage{tikz}
\usepackage{textcomp}
\usepackage{enumitem}

\usepackage[strings]{underscore}
\usepackage[left=1.62cm,right=1.62cm,top=0.7in]{geometry}
\usepackage{algorithm}
\usepackage{eurosym}

\begin{document}

\title{Market-Driven Flexibility Provision: A Tri-Level Optimization Approach for Carbon Reduction}

\author{
\IEEEauthorblockN{\textbf{Shijie Pan}\IEEEauthorrefmark{1}, \textbf{Gerrit Rolofs}\IEEEauthorrefmark{2}, \textbf{Luca Pontecorvi}\IEEEauthorrefmark{2}, \textbf{Charalambos Konstantinou}\IEEEauthorrefmark{1}}

\IEEEauthorblockA{\IEEEauthorrefmark{1}CEMSE Division, King Abdullah University of Science and Technology (KAUST)\\
\IEEEauthorrefmark{2}NEOM Energy and Water Company (“ENOWA”), NEOM, Saudi Arabia}

\IEEEauthorblockA{{
E-mail: \{firstname.lastname\}@kaust.edu.sa}, 
\{firstname.lastname\}@neom.com 
}}

\IEEEaftertitletext{\vspace{-1.2\baselineskip}}
\maketitle
\begin{abstract} 
The integration of renewable energy resources (RES) in the power grid can reduce carbon intensity, but also presents certain challenges. The uncertainty and intermittent nature of RES emphasize the need for flexibility in power systems. 
Moreover, there are noticeable mismatches between real-time electricity prices and carbon intensity patterns throughout the day. These discrepancies may lead customers to schedule energy-intensive tasks during the early hours of the day, a period characterized by lower electricity prices but higher carbon intensity. This paper introduces a novel and comprehensive framework aimed at encouraging customer participation in electricity markets and aligning their flexibility with carbon intensity trends. The proposed approach integrates an incentive-based tariff with a tri-level optimization model, where customers are motivated to submit flexibility bids and, in return, receive financial rewards based on their contributions. The tri-level model ensures a dynamic interaction between the market operation platform (MOP) and end-users. Simulations are performed on a modified IEEE-33 bus system, supported by two scenarios with different RES generations and customer behaviors. Results demonstrate the effectiveness of the proposed framework in guiding the customers' consumption behaviors towards low carbon intensity.
\end{abstract}
\vspace{-2mm}
\vspace{-2mm}

\section*{Nomenclature}
\addcontentsline{toc}{section}{Nomenclature}
\begin{IEEEdescription}[\IEEEusemathlabelsep\IEEEsetlabelwidth{$V_1,V_2,V_3$}]

\item[\textit{Sets and Indexes}] 
\item[$\mathcal{T}$] Set of hours in a day, indexed by $t$.
\item[$\mathcal{B}$] Set of buses, indexed by $i,j,k$.

\item[\textit{Variables}] 
\item[$P_{i,t}$] Tariff of bus $i$ at time $t$ (\$$/MWh$, \$$/MW$).
\item[$P^L_{i,t}$] Tariff of bus $i$ at time $t$ for lower consumption (\$$/MWh$, \$$/MW$).
\item[$P^M_{i,t}$] Tariff of bus $i$ at time $t$ for suitable consumption (\$$/MWh$, \$$/MW$).
\item[$P^U_{i,t}$] Tariff of bus $i$ at time $t$ for higher consumption (\$$/MWh$, \$$/MW$).

\item[$E^N_{i,t}$] Expected load demand on bus $i$ at time $t$ ($MW$).
\item[$E^A_{i,t}$] Actual load demand on bus $i$ at time $t$ ($MW$).
\item[$E^{BESS}_{i,t}$] BESS energy consumption/supply on bus $i$ at time $t$ ($MW$).
\item[$E^C_{t}$] Additional electricity bought from energy market at time $t$ ($MW$).
\item[$E^B_{i,t}$] Flexibility bid cleared on bus $i$ at time $t$ ($MW$).
\item[$E^L_{i,t}, E^U_{i,t}$] Upper and lower bounds of the tolerance range of energy consumption on bus $i$ at time $t$ ($MW$).
\item[$NL_{t}$] System net load at time $t$ ($MW$).
\item[$G_{i,t}$] PV power generation on bus $i$ at time $t$ ($MW$).
\item[$p^{ch}_{i,t},p^{dis}_{i,t}$] Charging and discharging energy of BESS on bus $i$ at time $t$ ($MW$).
\item[$p_{i,j,t},p_{j,k,t}$] Power flow from bus $i/j$ to bus $j/k$ at time $t$ ($MW$).
\item[$soc_{i,t}$] State of charge of BESS on bus $i$ at time $t$ ($MWh$).
\item[$d_{i,t}$] Variable representing upward/downward flexibility bid on bus $i$ at time $t$.

\item[\textit{Parameters}] 
\item[$\Delta t$] Time interval, hourly.
\item[$\delta$ ] Adjustable range of the flexible loads.
\item[$\epsilon$ ] Width of the optimal range in the incentive function.
\item[$\eta$] BESS charging and discharging efficiency.
\item[$\lambda_{i,t}$] Unit prices of flexibility bids on bus $i$ at time $t$ (\$$/MWh$, \$$/MW$).
\item[$\pi_{t}$] Unit prices of importing electricity at time $t$ (\$$/MWh$, \$$/MW$).
\item[$E^{Sch}_{i,t}$] Scheduled load demand on bus $i$ at time $t$ ($MW$).
\item[$\overline{FL_{i,t}},\underline{FL_{i,t}}$] Upper and lower bounds of the flexible load demand on bus $i$ at time $t$ ($MW$).
\item[$\overline{p^{ch}_{i,t}},\overline{p^{dis}_{i,t}}$] Charging and discharging limits of BESS on bus $i$ at time $t$ ($MW$).
\item[$\overline{soc_{i,t}},\underline{soc_{i,t}}$] Upper and lower bounds of the BESS state of charge on bus $i$ at time $t$ ($MWh$).

\end{IEEEdescription}
\section{Introduction}
\label{section: Introduction}

Reducing carbon emissions has become a key focus in the development and operation of power grids \cite{chen2023towards}, playing a crucial role in building sustainable energy systems and mitigating climate change. Aligned with this vision, NEOM (ENOWA\footnote{ENOWA is the energy, water, and hydrogen subsidiary of NEOM, a major smart city and sustainable development project in Saudi Arabia. ENOWA is responsible for building 100\% renewable energy infrastructure to power NEOM’s industries, businesses, and communities.}) \cite{NEOM} aims to establish the world’s first large-scale renewable energy system and climate-positive society. Upon completion, NEOM will rely 100\% on renewable energy sources (RES), including solar, wind and green hydrogen to meet its energy demands. The integration of these RES not only minimizes reliance on traditional fossil fuels but also significantly reduces carbon intensity.

Nevertheless, the intermittent and non-dispatchable nature of RES introduces significant challenges for maintaining balance within the energy system \cite{strbac2021decarbonization}. Ensuring grid stability and a reliable energy supply increasingly depends on the active participation of the demand side, either through distributed energy resources (DER) or by adapting energy consumption behaviors to provide flexibility \cite{d2022exploiting, aftab2025demand, 9340265}. Research in \cite{mohandes2020incentive} demonstrates that pricing models designed to incentivize demand-side flexibility can enhance customer participation, improving social welfare and minimizing RES curtailment. An incentive-based framework for flexibility provision in smart grids is proposed in \cite{hussain2023novel}. Experimental results show its capability to increase the monetary benefits of flexibility services and reduce peak load. However, its proposed incentive/penalty function requires customers to enforce the desired demand strictly, i.e., a slight excess of power consumption will be penalized, which reduces the customers' willingness to participate in the program.

\begin{figure}
    \centering
    \includegraphics[width=0.82\linewidth]{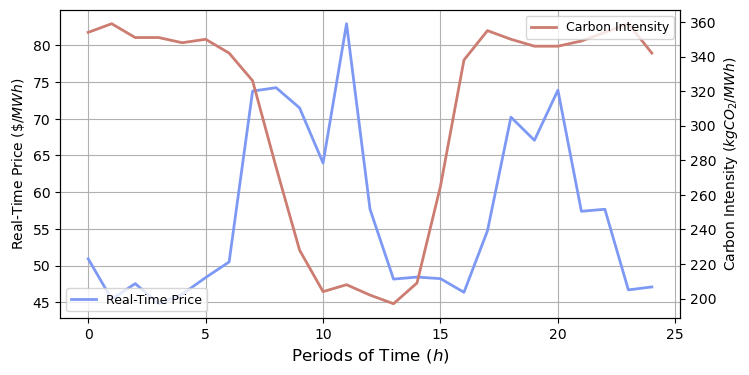}
    \vspace{-2mm}
    \caption{Real-time price \cite{RTPCali} and carbon intensity \cite{CarbonIntensityCali} of a typical day in California.}
    \label{fig:CarbonIntensity}
    \vspace{-5mm}
\end{figure}

In addition, from a carbon reduction perspective, there is a mismatch between the existing electricity prices and the carbon intensity profiles during the day \cite{10807052}. For instance, as shown in Fig. \ref{fig:CarbonIntensity}, real-time electricity prices in California peak during the two highest demand periods of the day, whereas carbon intensity reaches its lowest point at noon when solar PV generation is at its maximum due to abundant sunlight. The lower electricity prices at midnight motivate the customers to shift their load demand to these hours, helping reduce the peak demand but leading to an increase in carbon intensity. 

Therefore, it is crucial to develop approaches that not only incentivize customer participation in flexibility markets but also explicitly incorporate carbon intensity reduction as a key objective. This work examines the impact of incentive-based tariffs on enhancing system flexibility, particularly in the context of 100\% RES  integration. We propose a framework that enables flexibility prosumers to actively participate in the flexibility market by submitting bids and receiving financial incentives. The main contributions of this paper are as follows:
\begin{enumerate}
    \item An incentive-based tariff is designed to encourage customers to provide flexibility. Compared with existing works, the proposed incentive function offers users tolerance margins, balancing encouraging participation and probable overly sensitive penalties that could discourage participation. 
    
    \item A novel framework is proposed to design tariffs dynamically based on daily carbon intensity patterns. Customers are incentivized to participate in the flexibility market through bidding mechanisms in the day-ahead and real-time markets, promoting active engagement and alignment with carbon reduction objectives.
    
    \item Simulation studies are conducted on a modified IEEE 33-bus system. Two cases with different RES generations and load profiles are considered. The results demonstrate the effectiveness of the designed incentive function and flexibility provision framework.
\end{enumerate}

The rest of the paper is as follows. 
Section~\ref{section: Flexibility Market} gives a short introduction to the flexibility market. Section~\ref{section: Methodology} presents the proposed incentive-based tariff and flexibility provision framework. 
Simulation results are presented in Section~\ref{section: Simulation}, while 
Section~\ref{section: Conclusion} concludes the work.

\vspace{-1mm}
\section{NEOM's Flexibility Market Mechanism}
\label{section: Flexibility Market}
NEOM aims to develop a flexible and resilient energy market tailored to its unique challenges, including high renewable energy penetration, variability in supply and demand, and the need for efficient grid balancing. The flexibility market will serve as an essential mechanism to address these challenges by enabling real-time supply-demand balancing. Market operators should oversee both day-ahead and intra-day operations, ensuring efficient resource utilization. Acting as a unified platform, the market optimizes the deployment of balancing resources and facilitates the commercialization of flexibility services in a competitive environment. This structure allows for the selection of the most effective solutions to maintain system stability and reliability during real-time operations.

To encourage diverse forms of flexibility, NEOM's flexibility market creates an open and transparent environment where stakeholders can offer their services, allowing for comparison and selection of the best solutions by the market operator. 
Furthermore, the flexibility market offers clear price signals that guide investment decisions in various types of flexibility assets, whether from ENOWA or third-party providers, including demand response, electricity storage, and distributed energy generation. This design not only fosters innovation and efficiency but also aligns with global standards, supporting third-party engagement and investment.

As the central communication and information-sharing hub between the market operator and participants in the flexibility market, the Market Operation Platform (MOP) plays a pivotal role. It serves as the main source of data for balancing and control purposes, consolidating all necessary information for optimization. Through the MOP, active market participants can submit their nominations and flexibility bids, while also providing a way for other flexibility providers and contractors to contribute to system balancing and control. Moreover, the MOP serves as the platform for disseminating the results of system-wise optimization to market participants.

\vspace{-1mm}
\section{Methodology}
\label{section: Methodology}

\subsection{Incentive-Based Tariff Design}
\begin{figure}
    \centering
    \includegraphics[width=0.82\linewidth]{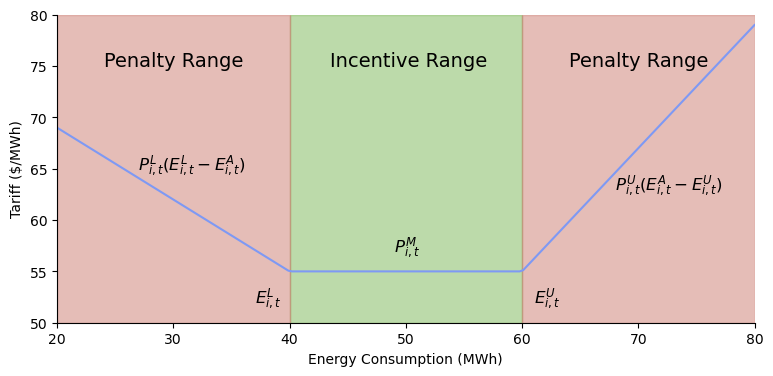}
    \vspace{-2mm}
    \caption{An illustrative example of the incentive-based tariff.}
    \label{fig:Tariff}
    \vspace{-3mm}
\end{figure}

Research has shown that electricity prices influence customer behavior \cite{nakabi2019ann}. In this work, we design an incentive-based tariff to encourage customers to maintain their energy consumption within a dynamic optimal range, denoted as $[E^L_{i,t}, E^U_{i,t}]$. The tariff structure is defined as follows: \begin{equation}\label{Eq: Tariff}
    \begin{aligned}
    P_{i,t} = &P^M_{i,t} 
 + P^L_{i,t} \max\{E^A_{i,t}-E^U_{i,t},0\} 
    \\&+ P^U_{i,t} \max\{E^L_{i,t}-E^A_{i,t},0\}
    \end{aligned}
\end{equation}
Eq. \eqref{Eq: Tariff} can be rewritten as:
\begin{equation}\label{Eq: Tariff1}
    P_{i,t} = \left\{\begin{aligned}
    &P^L_{i,t}(E^L_{i,t}-E^A_{i,t})+P^M_{i,t}, &E^A_{i,t}<E^L_{i,t}\\
    &P^M_{i,t}, &E^L_{i,t}\le E^A_{i,t}\le E^U_{i,t}\\
    &P^U_{i,t}(E^A_{i,t}-E^U_{i,t})+P^M_{i,t}, &E^A_{i,t}> E^U_{i,t}
    \end{aligned}\right..
\end{equation}
The tariff model, illustrated in Fig. \ref{fig:Tariff}, incentivizes customers to maintain their consumption within the optimal range $[E^L_{i,t}, E^U_{i,t}]$ by applying a base rate $P^M_{i,t}$. Any deviation beyond this range incurs a penalty proportional to the degree of deviation, promoting load flexibility and aligning energy usage with system-level objectives. 
Unlike the approach in \cite{hussain2023novel}, which enforces strict penalties for minor deviations, our model allows customers to adjust their consumption within $[E^L_{i,t}, E^U_{i,t}]$ without financial consequences. This flexibility prevents customer dissatisfaction and encourages participation in demand-side management programs. 
Furthermore, the bidirectional penalty structure not only discourages excessive consumption but also aligns energy usage with carbon intensity patterns. It promotes lower consumption during high-carbon-intensity periods in the morning and evening while encouraging increased usage during midday when renewable generation is abundant.

\vspace{-2mm}
\subsection{Flexibility Provision Framework Design}
\begin{figure}
    \centering
    \includegraphics[width=0.6\linewidth]{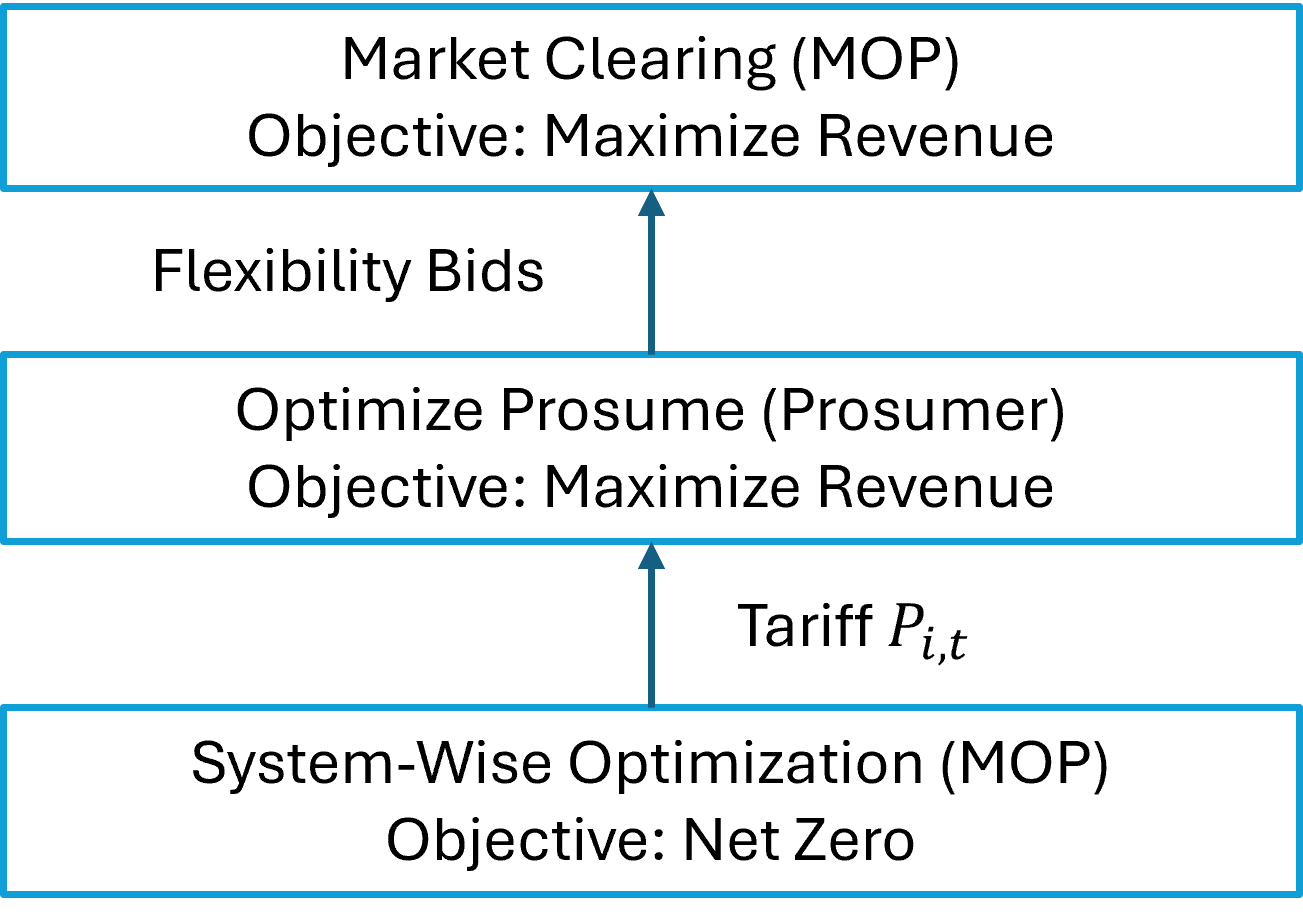}
    \vspace{-2mm}
    \caption{The tri-level optimization for maximizing the flexibility revenues.}
    \vspace{-4mm}
    \label{fig:TriLevel}
\end{figure}

In order to maximize both the revenues and social welfare for customers, we propose a tri-level optimization framework, as illustrated in Fig.~\ref{fig:TriLevel}. As discussed in \cite{deangelo2021energy}, achieving net-zero scenarios requires a significant reduction in the carbon intensity of final energy consumption. Accordingly, at the lower level, the MOP performs a system-wide optimization to achieve net-zero, aligning with the requirements of 100\% RES integration while also reducing carbon intensity. The solutions from this lower-level optimization provide the MOP with the expected load demand $E^N_{i,t}$ and the tolerance range $[E^L_{i,t}, E^U_{i,t}]$, which are then incorporated into the flexibility tariff $P_{i,t}$. At the middle level, flexibility users optimize their energy consumption according to the received tariff, generating flexibility bids that maximize their revenue. Finally, at the upper level, the MOP conducts market clearing with the received bids, focusing on maximizing overall revenue.

\textbf{Lower-level problem}:
The objective of the lower-level optimization is to achieve net zero under a 100\% RES generation scenario. The problem can be formulated as:
\begin{flalign} \label{LL}
&\mathop{minimize}\limits_{E^N_{i,t},E^L_{i,t},E^U_{i,t}} \sum_{t\in \mathcal{T}}|NL_t| \\
\textrm{s.t.} \quad & NL_t = \sum_{i\in \mathcal{B}}G_{i,t}-\sum_{i\in \mathcal{B}}E^N_{i,t} \label{LL1}\\
& (1-\delta)E^{Sch}_{i,t} \leq E^N_{i,t} \leq (1+\delta)E^{Sch}_{i,t} \label{LL2}\\
& E^L_{i,t} = (1-\epsilon) min\{E^{Sch}_{i,t}, E^N_{i,t}\} \label{LL3}\\
& E^U_{i,t} = (1+\epsilon) max\{E^{Sch}_{i,t}, E^N_{i,t}\} \label{LL4}\\
& E^{BESS}_{i,t} = p^{ch}_{i,t}-p^{dis}_{i,t} \label{LL5}\\
& 0 \leq p^{ch}_{i,t} \leq \overline{p^{ch}_{i}} \label{LL6}\\
& 0 \leq p^{dis}_{i,t} \leq \overline{p^{dis}_{i}} \label{LL7}\\
& p^{ch}_{i,t}\cdot p^{dis}_{i,t} = 0 \label{LL8}\\
& \underline{soc_{i,t}} \leq soc_{i,t} \leq \overline{soc_{i,t}} \label{LL9}\\
& soc_{i,t+1} = soc_{i,t} + \Delta t(\eta_i p^{ch}_{i,t}-\frac{p^{dis}_{i,t}}{\eta_i}) \label{LL10}\\
& \sum p_{j,k,t} = p_{i,j,t}+G_{j,t}-E^N_{j,t}-E^{BESS}_{j,t} \label{LL11}.
\end{flalign}
\eqref{LL1} calculates the system's total net load. \eqref{LL2} represents the flexible loads (FLs) model. \eqref{LL3} and \eqref{LL4} imply that the designed tariff range $[E^L_{i,t}, E^U_{i,t}]$ should contain the scheduled demand $E^{sch}_{i,t}$, preventing customers from incurring losses when participating in flexibility incentive programs. \eqref{LL5}-\eqref{LL10} define the battery energy storage system (BESS) model, and \eqref{LL11} calculates the active power flow.

\textbf{Middle-level problem}:
The objective of the flexibility prosumers is to maximize their revenues by utilizing the available flexibility. The revenues consider both the cost of tariff and also the profits that the prosumers can gain with the corresponding energy consumption \cite{di2018optimal}. $f(E^A_{i,t})$ in \eqref{ML} is a utility function that indicates the relationship between profits and energy consumption, which can be linear, exponential, or other form that fits the consumption behaviors. The optimization problem can be formulated as:
\begin{flalign} \label{ML}
\sum_{i\in \mathcal{B}}&\mathop{maximize}\limits_{E^A_{i,t}} \sum_{t\in \mathcal{T}}f(E^A_{i,t})-P_{i,t} E^A_{i,t} \\
\textrm{s.t.} \quad &\underline{FL_{i,t}} \leq E^A_{i,t} \leq \overline{FL_{i,t}} \label{ML1} \\ 
&P_{i,t} = P^L_{i,t} \max\{E^A_{i,t}-E^U_{i,t},0\} \notag \\
&+P^U_{i,t} \max\{E^L_{i,t}-E^A_{i,t},0\}+P^M_{i,t}\label{ML2}.
\end{flalign}
\eqref{ML1} bounds the consumption range that can be adjusted based on the FL model. \eqref{ML2} calculates the incentive tariff for the customer.

\textbf{Upper-level problem}: The upper-level problem clears the markets. It follows the formulation of the day-ahead and reserve markets in Eq. (6c) from \cite{paredes2023stacking}: 
\begin{flalign} \label{UL}
&\mathop{maximize}\limits_{E^B_{i,t}}\sum_{t\in \mathcal{T}}\sum_{i\in \mathcal{B}}P_{i,t} E^A_{i,t}-\lambda_{i,t}E^B_{i,t}-\pi_{t}E^C_{t}\\
\textrm{s.t.} \quad & \sum_{i\in \mathcal{B}}G_{i,t} + \sum_{i\in \mathcal{B}}E^{BESS}_{i,t}+E^C_{t} -\sum_{i\in \mathcal{B}}(E^{Sch}_{i,t}+d_{i,t} E^B_{i,t}) \geq 0 \label{UL1}\\
&0 \leq E^B_{i,t} \leq |E^A_{i,t}-E^{Sch}_{i,t}| \label{UL2}\\
& d_{i,t} = \frac{|E^A_{i,t}-E^{Sch}_{i,t}|}{(E^A_{i,t}-E^{Sch}_{i,t})} \label{UL3}.
\end{flalign}
The revenue considers the actual power consumption of the customers, the cost of clearing flexibility bids, and the cost of importing power. \eqref{UL1} ensures the power supply in the system. If power generation is insufficient, additional capacities $E^C_{t}$ are imported. \eqref{UL2} sets out the scope of clearing bids. \eqref{UL3} defines the sign of $E^B_{i,t}$ in \eqref{UL1}.

\textbf{Real-time bidding and clearing}:
If the number of flexibility bids exceeds the requirement, the unaccepted bids at this stage will be prioritized in real-time flexibility trading.

\begin{figure}
    \centering
    \includegraphics[width=1\linewidth]{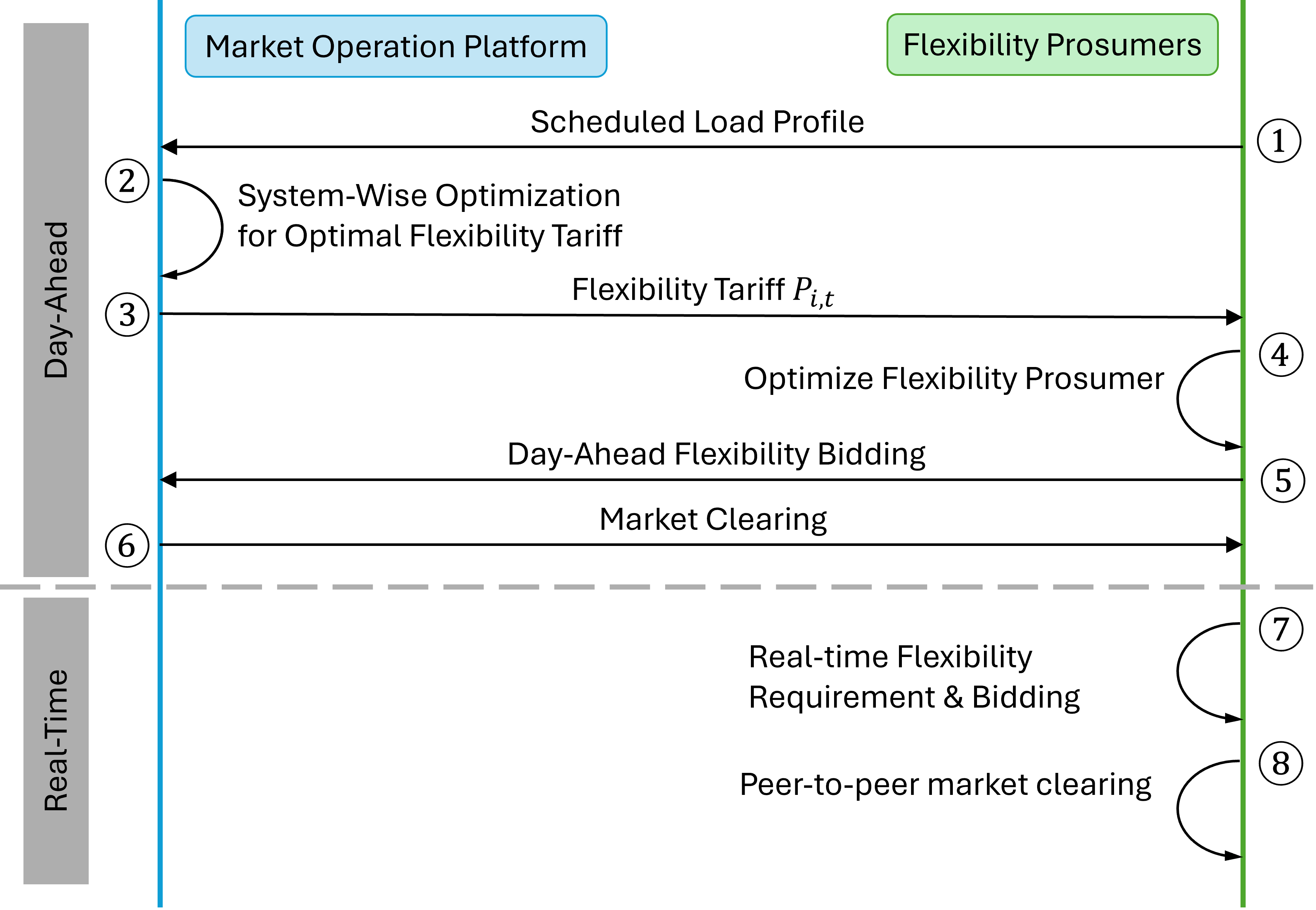}
    \vspace{-5mm}
    \caption{Timeline diagram of the market operation platform (MOP) interactions with flexibility prosumers.}
    \vspace{-1mm}
    \label{fig:Timeline}
\end{figure}

The interaction between the MOP and flexibility prosumers follows the timeline outlined in Fig.~\ref{fig:Timeline}.  
In the first step, the flexibility prosumers send their scheduled or forecasted load profile $E^F_{i,t}$ to the MOP. In step 2, the MOP performs a system-wise optimization based on the load and generation profile for the next day, aiming to achieve net-zero and reduce carbon intensity. The optimization provides the optimal expected load profiles $E^N_{i,t}$ along with the tolerance range $[E^L_{i,t}, E^U_{i,t}]$. Using the results from the optimization, the MOP determines and sends the tariff to the flexibility prosumers in step 3. Then, in step 4, the flexibility prosumers optimize and reschedule their energy consumption based on the given tariff and generate flexibility bids. The bids are then sent to the MOP in step 5, and the MOP performs market clearing in step 6. Steps 1 through 6 are executed a day ahead to ensure optimal flexibility scheduling. In real-time operations, if there are any remaining day-ahead bids, they are cleared first in step 7 when flexibility requirements arise. Additional requirements are cleared peer-to-peer in step 8 \cite{khorasany2022framework}.

\vspace{-2mm}
\section{Simulation Results}
\label{section: Simulation}

\begin{figure}
    \centering
    \includegraphics[width=0.99\linewidth]{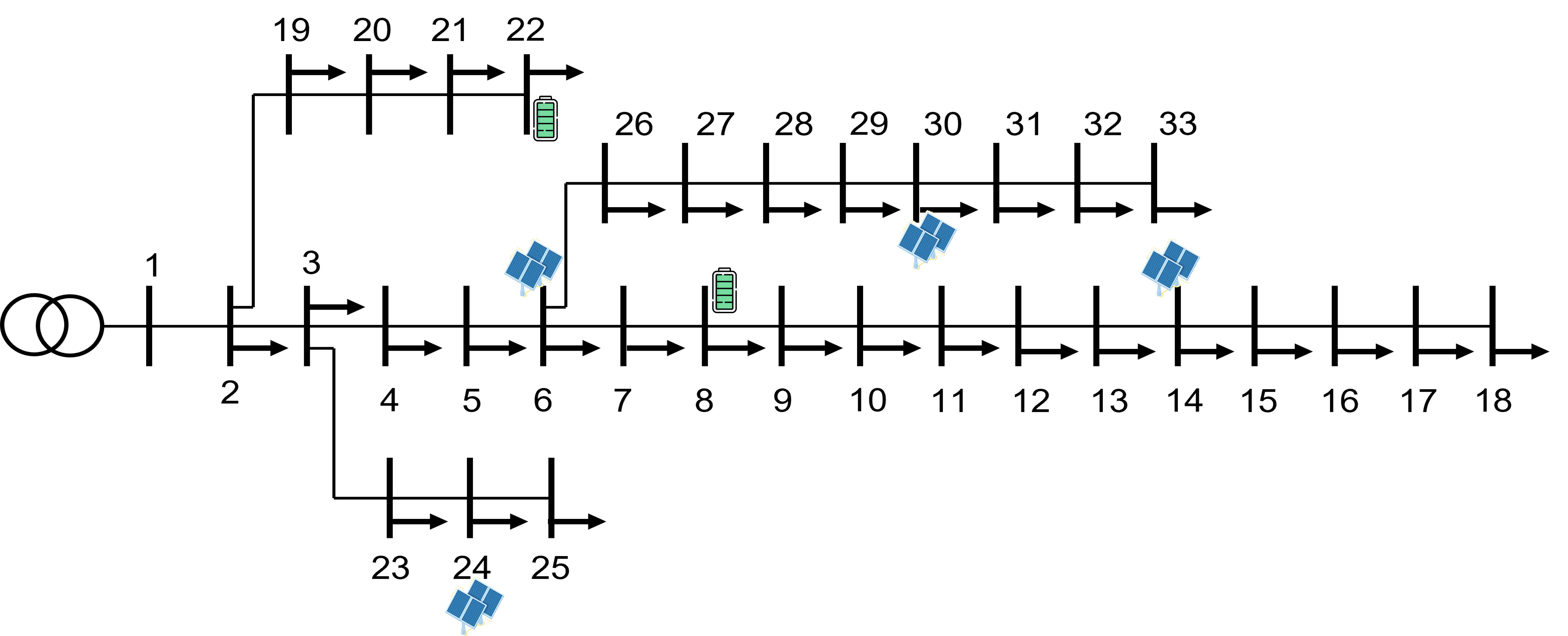}
    \vspace{-6mm}
    \caption{The modified IEEE 33-bus test system.}
    \label{fig:IEEE33M}
    \vspace{-3mm}
\end{figure}

\begin{figure}
    \centering
    \vspace{-5mm}
    \includegraphics[width=0.97\linewidth]{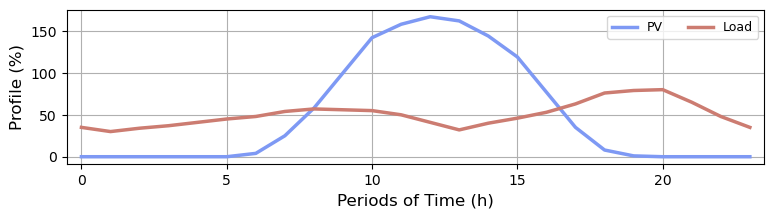}
    \vspace{-3mm}
    \caption{The 24-h PV and load profile.}
    \vspace{-3mm}
    \label{fig:PVLoadProfile}
\end{figure}

A modified IEEE 33-bus system, as depicted in Fig. \ref{fig:IEEE33M}, is used in the experiments to validate the proposed methodology. 4 PVs and 2 BESSs are considered in the system, and their locations are shown in the figure. The capacities of the BESSs are both $1.5$ MW, and the capacities of the PVs are [$1.2$, $0.715$, $1.2$, $0.6$] MW, based on the study in \cite{pan2023optimal}. The 24-h PV and load profiles are shown in Fig. \ref{fig:PVLoadProfile}. $P^L, P^U$ are set to $1$, with $\delta=0.2, \eta=0.95, \epsilon=0.05$. 

\begin{figure}
    \centering
    \includegraphics[width=0.95\linewidth]{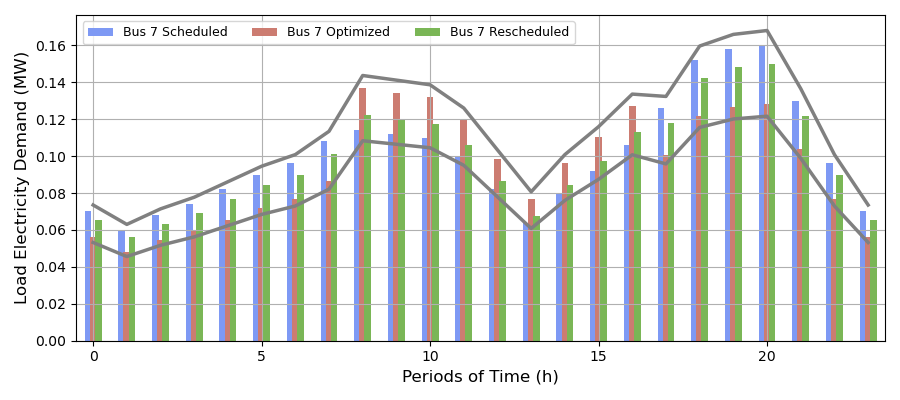}
    \vspace{-3mm}
    \caption{The 24-h load demands on bus 7.}
    \vspace{-4mm}
    \label{fig:Bus7Respond}
\end{figure}

Fig. \ref{fig:Bus7Respond} illustrates the 24-hour demand profile of a flexibility prosumer at bus 7 on a sample day. The blue bars represent the customer's scheduled consumption before any optimizations. The red bars indicate the expected consumption values, as determined by the system-wise optimization, which are higher during the day due to the abundant generation from PV generations. At night, the expected consumption decreases as the customer is encouraged to reduce their demand in order to minimize carbon intensity. The gray curves, which represent the $E^L_{7,t}$ and $E^U_{7,t}$ in the designed incentive-based tariff, are calculated based on the lower-level optimization results to motivate the customers to adjust their consumption behavior. It can be observed that the scheduled demands remain within the incentive range, ensuring that customers' interests are not negatively impacted by their inability to offer flexibility. In response to this tariff, the customer formulates an alternative consumption plan, depicted by the green bars, which reduces its demand at night and increases it during the day. With this plan, the customer is able to reserve flexibility (the difference between the green and blue bars) in advance and can participate in the flexibility market by submitting bids.

\begin{figure}
    \centering
    \includegraphics[width=0.95\linewidth]{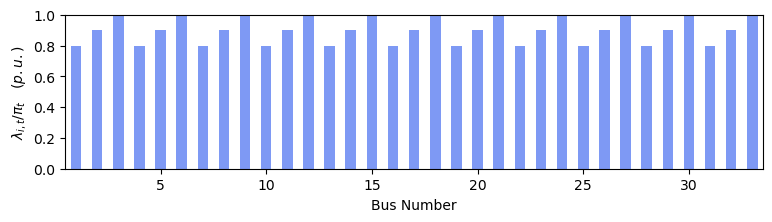}
    \vspace{-4mm}
    \caption{Unit prices of flexibility bids from prosumers at different buses.}
    \label{fig:BidPrice}
    \vspace{-4mm}
\end{figure}

After receiving the flexibility bids from the prosumers in the system, the MOP clears the market based on the prices of the bids and the cost of importing electricity. In the experiment, the unit prices of flexibility bids $\lambda_{i,t}$ are set proportionally based on the cost of importing electricity $\pi_{t}$ \cite{ruwaida2022tso}, as shown in Fig. \ref{fig:BidPrice}.

\begin{figure}
    \centering
    \vspace{-3mm}
    \includegraphics[width=0.95\linewidth]{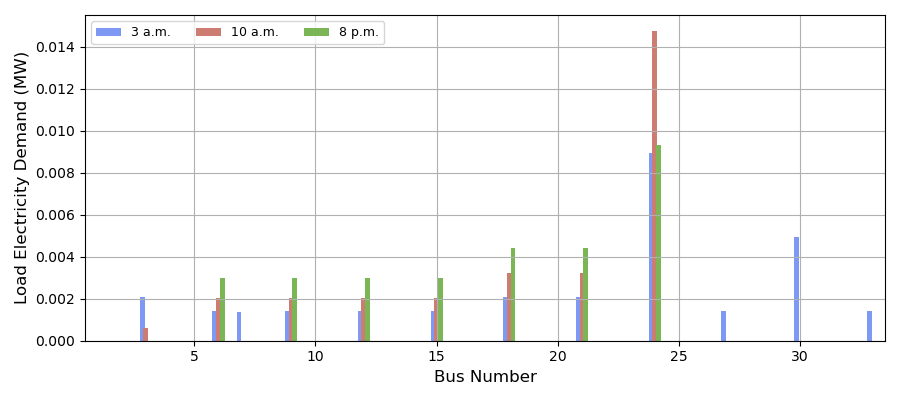}
    \vspace{-4mm}
    \caption{Bids cleared day-ahead on all system buses at different times.}
    \label{fig:DABiddings}
    \vspace{-4mm}
\end{figure}

\begin{figure}
    \centering
    \includegraphics[width=0.95\linewidth]{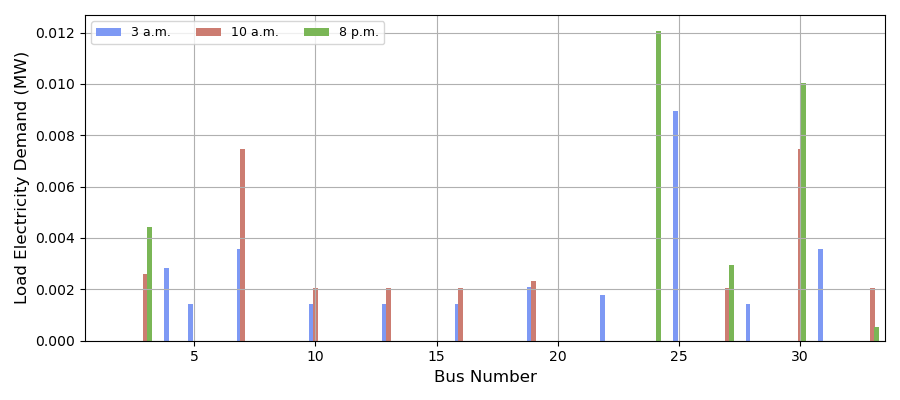}
    \vspace{-3mm}
    \caption{Bids cleared in real-time on all system buses at different times.}
    \label{fig:RTBiddings}
    \vspace{-3mm}
\end{figure}

\begin{figure}
    \centering
    \vspace{-5mm}
    \includegraphics[width=0.95\linewidth]{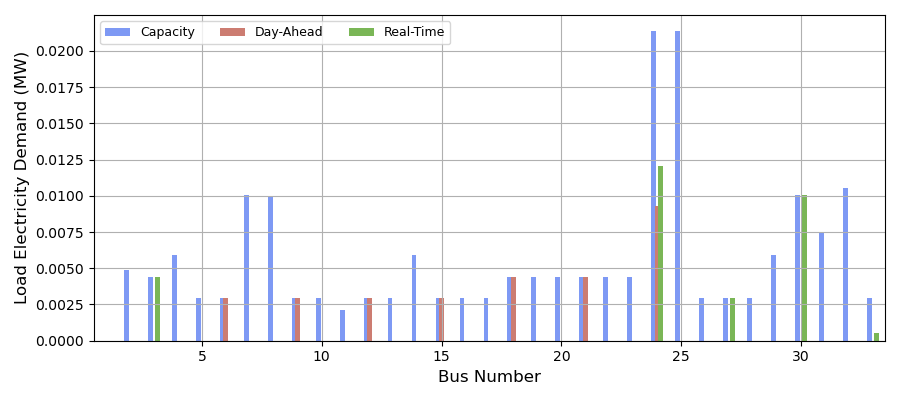}
    \vspace{-4mm}
    \caption{Flexibility capacity and bids cleared on different buses at 8 p.m.}
    \label{fig:Biddings}
    \vspace{-2mm}
\end{figure}

Fig. \ref{fig:DABiddings} presents the results of the day-ahead cleared bids. The figure demonstrates that the bids are cleared sequentially, with buses offering the lowest prices as presented in Fig. \ref{fig:BidPrice}. At 3 a.m., the bid from bus 7 is cleared due to the limited flexibility capacity of buses with lower bid prices. The remaining bids are held back and prioritized for clearing when flexibility requirements arise in real-time operations. Fig. \ref{fig:RTBiddings} shows the results of the bids that are cleared when there is a $30$ kW flexibility requirement in real-time. It can be observed that the remaining bids with lower prices are selected and cleared. The detailed relationship between the flexibility capacities and bid clearing process is illustrated in Fig. \ref{fig:Biddings}. 

\begin{figure}
    \centering
    \includegraphics[width=0.95\linewidth]{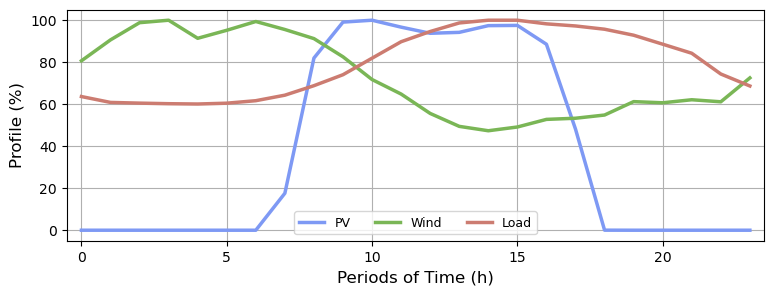}
    \vspace{-3mm}
    \caption{Generation \& load profile based on ENOWA data.}
    \vspace{-4mm}
    \label{fig:ProfileENOWA}
\end{figure}

\begin{figure}
    \centering
    \includegraphics[width=0.95\linewidth]{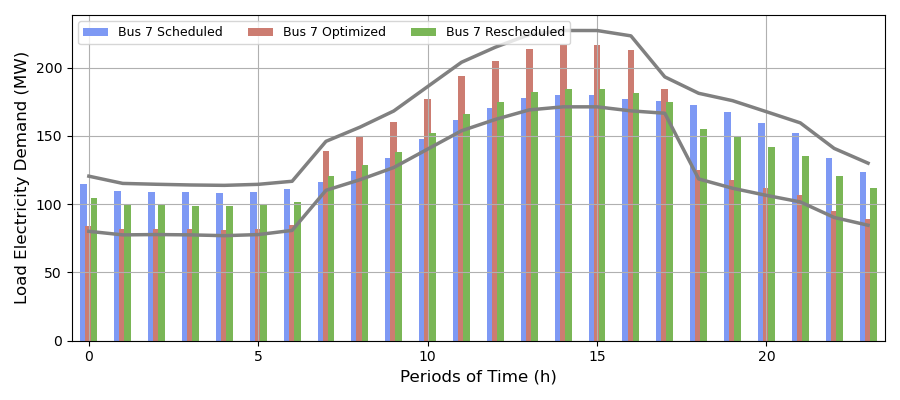}
    \vspace{-3mm}
    \caption{24-h load demands on bus $7$ based on ENOWA data.}
    \vspace{-4mm}
    \label{fig:ResponseENOWA}
\end{figure}

\begin{figure}
    \centering
    \includegraphics[width=0.95\linewidth]{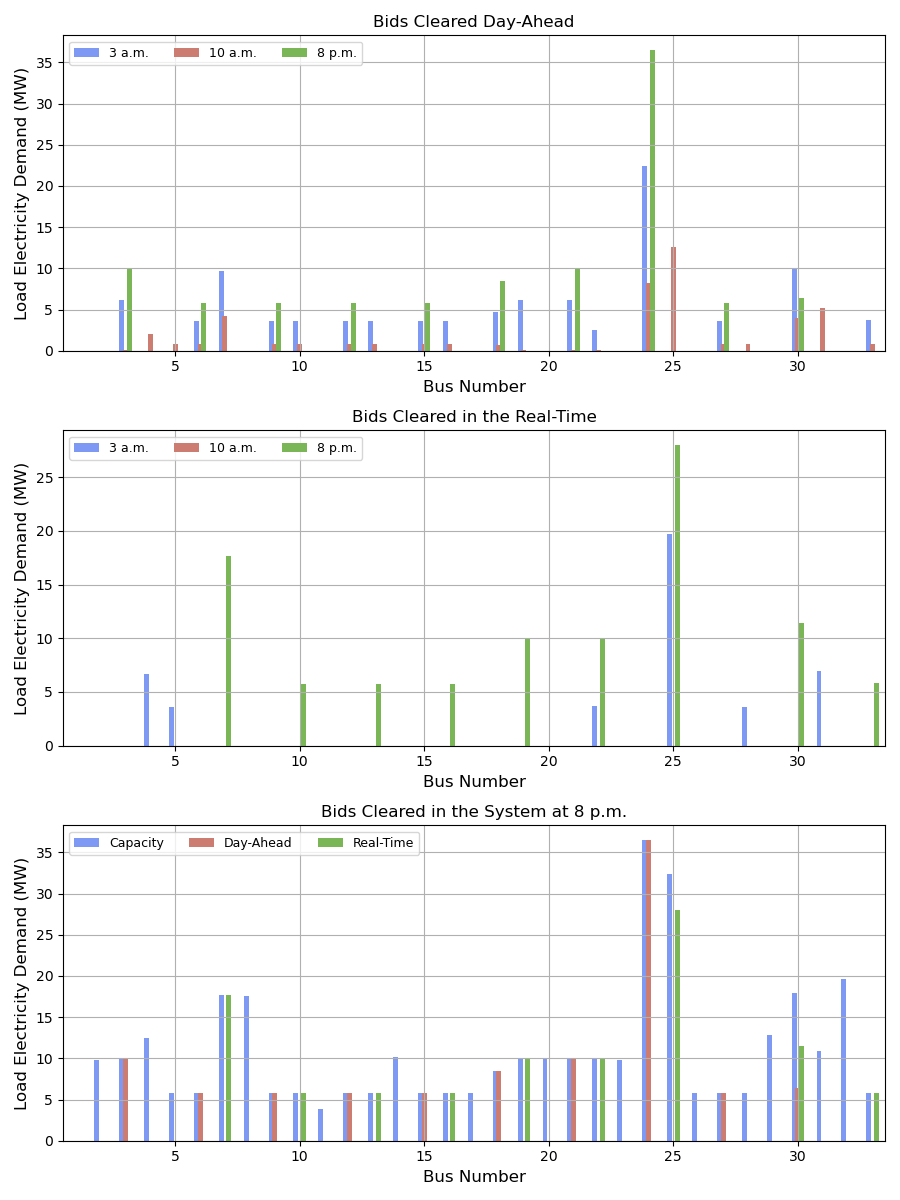}
    \vspace{-3mm}
    \caption{MOP market clearing results based on ENOWA data.}
    \vspace{-4mm}
    \label{fig:BiddingsENOWA}
\end{figure}

The proposed framework is further examined in an additional scenario that incorporates wind generations, aligning with ENOWA's future development plans. The consumers also exhibit different energy consumption behaviors, characterized by peak demand occurring during daytime hours. Fig. \ref{fig:ProfileENOWA} illustrates the RES generation and load profile of the scenario. The prosumer's flexibility response at bus 7 is presented in Fig. \ref{fig:ResponseENOWA}. As shown, the implementation of the incentive tariff leads to an increase in daytime power consumption among consumers while encouraging a reduction in electricity usage at night, thereby contributing to a decrease in carbon intensity. Fig. \ref{fig:BiddingsENOWA} presents the results of flexibility market clearing, showing the connections between day-ahead and real-time flexibility bidding. It can be observed that due to the high flexibility requirement day ahead, many bids are cleared, though they do not have the lowest price. At 8 p.m., the bids are cleared sequentially in real-time due to sufficient capacity. However, at 3 a.m., a bid at bus 4 is cleared due to insufficient remaining bids after day-ahead market clearing.

\vspace{-2mm}
\section{Conclusion}
\label{section: Conclusion}

In this study, an incentive-based tariff is formulated to encourage the customers in the power system to provide flexibility. The proposed flexibility provision framework incorporates the consideration of reducing carbon intensity. A tri-level optimization is designed to maximize the revenues for both, the market operator and the flexibility prosumers. By interacting with the MOP through receiving the incentive tariff and sending the flexibility bids, the customers are able to earn rewards and contribute to the power system flexibility. Results on two different scenarios demonstrate that the presented methodology can guide the customers' behavior and contribute to decreased carbon intensity. The proposed framework provides a solution for enhancing power system flexibility and aligning monetary incentives with reducing carbon intensity.
\vspace{-1mm}
\section*{{Acknowledgment}}
{This paper is an outcome of a larger 2-year project to develop intelligent solutions for grid technologies for ENOWA (NEOM) energy systems funded by ENOWA (NEOM) through
a technical consultancy agreement with KAUST.}
\vspace{-2mm}
\bibliographystyle{IEEEtran}
\bibliography{refs}
\end{document}